\documentclass[twocolumn,showpacs,preprintnumbers,amsmath,amssymb,aps]{revtex4}
\usepackage[pdftex]{graphicx}
\usepackage{bm}

\begin{document}
\title{Temperature-dependent scanning tunneling spectroscopy on the Si(557)-Au surface}
\author{M. Sauter$^1$}
\author{R. Hoffmann$^1$}
\author{C. S\"urgers$^1$}
\email{christoph.suergers@kit.edu}
\author{H. v. L\"ohneysen$^{1,2}$}
\affiliation{$^1$Physikalisches Institut and DFG Center for Functional
Nanostructures (CFN), Karlsruhe Institute of Technology, P.O. Box 6980, 76049 Karlsruhe, Germany\\
$^2$Institut f\"ur Festk\"orperphysik, Karlsruhe Institute of Technology, P.O. Box 3640, 76021 Karlsruhe, Germany}

\date{\today}

\begin{abstract}
Room-temperature and low-temperature (77 K) scanning tunneling spectroscopy (STS) and voltage-dependent scanning tunneling microscopy (STM) data are used to study the local electronic properties of the quasi-one-dimensional Si(557)-Au surface in real space. A gapped local electron density of states near the $\bar{\Gamma}$ point is observed at different positions of the surface, i.e., at protrusions arising from Si adatoms and step-edge atoms. Within the gap region, two distinct peaks are observed on the chain of localized protrusions attributed to Si adatoms. The energy gap widens on both types of protrusions after cooling from room temperature to $T =$ 77 K. The temperature dependence of the local electronic properties can therefore not be attributed to a Peierls transition occurring for the step edge only. We suggest that more attention should be paid to finite-size effects on the one-dimensional segments.
\end{abstract}

\maketitle
\section{Introduction}
Metal-induced one-dimensional (1D) reconstructions of vicinal semiconductor surfaces are well suited for studying the electronic properties of 1D systems. For instance, submonolayer deposition of gold on Si(111) substrates with an intentional miscut allows the preparation of 1D surface structures with taylored electronic states by tuning the tilt angle of the vicinal surface \cite{BarkeSSC,Crain04,Snijders2010}. In these quasi-1D electronic systems, the electron-electron and electron-phonon interactions play important roles. While a Luttinger liquid is formed due to the electron-electron interaction, in, e.g., carbon nanotubes \cite{Gao04}, the electron-phonon interaction gives rise to a Peierls instability and a charge-density-wave ground state as found in various inorganic and organic quasi-1D compounds \cite{Kagoshima,Gruener,Carpinelli,Aruga}. However, in strictly 1D systems large fluctuations of the order parameter prevent a Peierls (second-order) phase transition at finite temperatures. For $T > 0$ a pseudo energy-gap is then formed in the electronic density of states due to short-range correlations. On Si(553)-Au obtained on Si(111) substrates with an intentional miscut of $12.5^{\circ}$ towards [\=1\=12], multiple distortions of the surface lattice have been observed by STM and have been interpreted in terms of Peierls distortions with different and competing periodicities \cite{Ahn05,Snijders}. 

The Si(557)-Au surface, forming on Si(111) with a miscut angle of $9.5^{\circ}$ towards [\=1\=12], has been proposed to contain one atomic Au row and an atomic step per surface unit-cell \cite{Losio01}. This surface has attracted considerable interest due to its unusual electronic properties, such as two closely separated occupied bands S$_1$ and S$_2$ just below the Fermi level $E_F$ observed by angular resolved photoemission electron spectroscopy (ARPES) \cite{Losio01,Segovia99,Ahn03}. While first experiments reported only a slight separation between those bands and a vanishing density of states (DOS) at the Fermi level which was interpreted as an indication of spin-charge separation in a Luttinger liquid \cite{Segovia99}, a later study resolved a larger separation and a finite DOS at the Fermi level \cite{Losio01}. From first-principles calculations of the surface band structure, the splitting of two proximate bands evolving from the Brillouin zone boundary was attributed to the large spin-orbit coupling due to the heavy Au at the surface \cite{Sanchez04}. Recently, experimental evidence for a splitting induced by spin-orbit interaction (Rashba effect) was reported to occur in one-dimensional chains on Si(553)-Au which exibits a similar doublet of bands as Si(557)-Au \cite{BarkePRL}. 

A combined photoemission and STM study has shown that the position of the low-energy band S$_2$ is independent of temperature whereas the position of the upper band $S_1$ shifts from the Fermi level $E_F$ at room temperature (RT) to below $E_F$ with decreasing temperature \cite{Ahn03}. This was interpreted as a metal-insulator transition in the upper band S$_1$ - associated with the step edge - due to a Peierls-like instability with a characteristic transition temperature $T_c \approx$ 270~K. The presence of a Peierls transition was supported by the observation of a dimerization in STM images at low temperature (LT) $T$ = 77 K which seemed to be absent at RT \cite{Ahn03,Yeom05}. However, previously reported well-resolved STM images detected such a dimerization also at RT, even without thermal fluctuation of the buckling \cite{Sauter07,Han2009}, putting the existence of a Peierls transition on this surface into question. Furthermore, an ARPES study using synchrotron radiation revealed that both bands $S_1$ and $S_2$ remain metallic for 83 K $\le T \le$\, 300 K, thus excluding the presence of a Peierls-type metal-insulator transition, at least above 83 K \cite{Kim2009}.

We have now investigated in detail the atomically resolved electronic structure of the Si(557)-Au surface with STS at RT and LT. We complement these measurements with new highly resolved STM images. Tunneling spectra obtained on both characteristic positions at the surface (step edge and adatoms) exhibit a gapped local electron density of states (LDOS) near the $\bar{\Gamma}$ point at 300 K and 77 K, in disagreement with what is expected from band-structure calculations \cite{Sanchez04}. 

\section{Experimental}
\begin{figure}
\includegraphics[width=0.7\columnwidth]{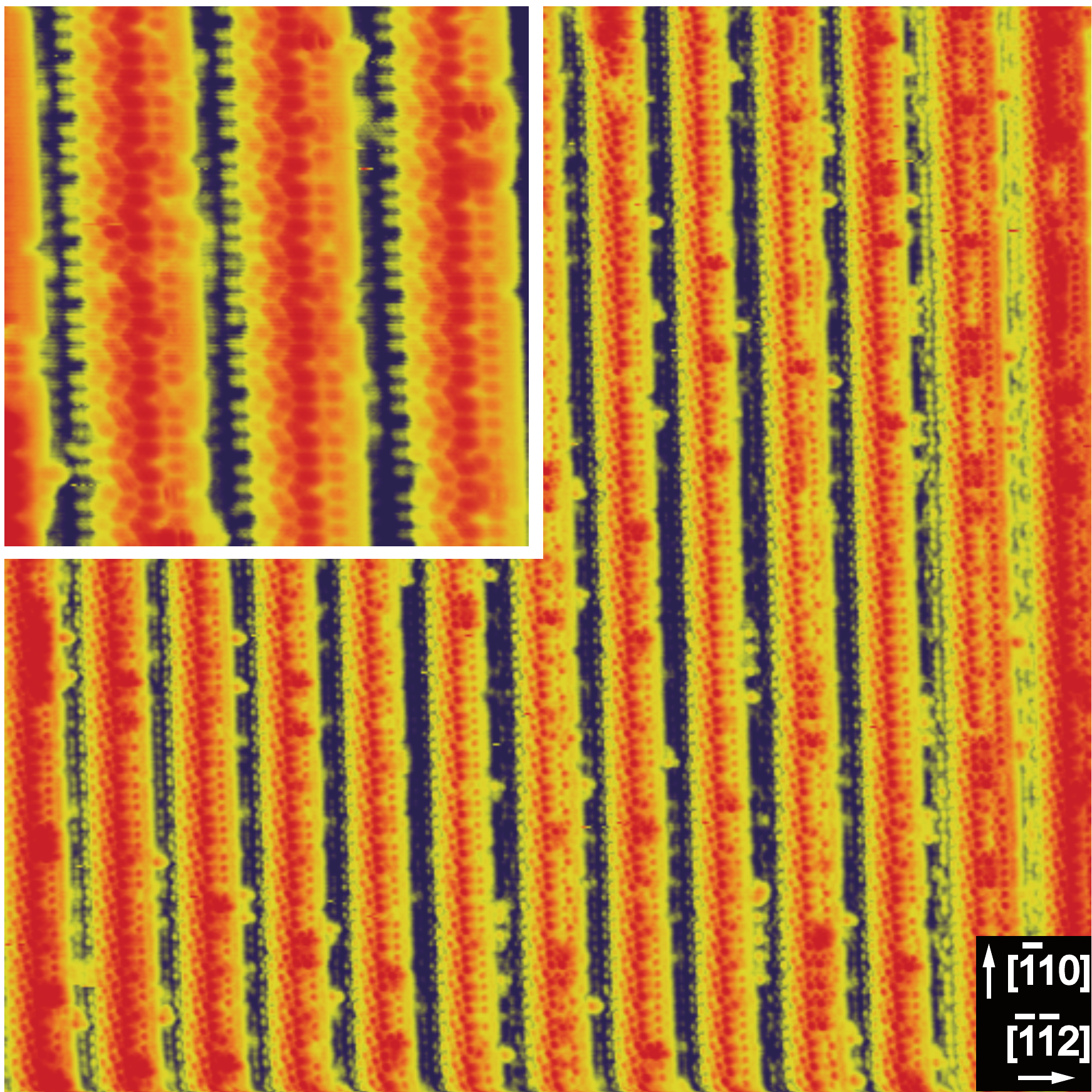}
\caption[]{STM image (80 nm $\times$ 80 nm) of the vicinal Si(111) surface after annealing and before Au deposition, taken at a sample voltage $U = +2$ V (tunneling current $I$ = 0.1 nA) at room temperature. Inset shows an image (20 nm $\times$ 20 nm) obtained with higher spatial resolution.} 
\label{fig0}
\end{figure}

The Si(557)-Au surface was obtained by first annealing a Si(111) substrate with an intentional miscut of $10^{\circ} \pm 0.5^{\circ}$ towards [\=1\=12] in ultra-high vacuum \cite{Schoeck03,Schoeck06}. After annealing, the STM images revealed a stepped surface structure of Si(111)-7$\times$7 akin to the [7 7 10] orientation of vicinal Si(111) \cite{Teys}, see Fig. \ref{fig0}. The formation of the quasi-1D Si(557)-Au surface was induced by depositing 0.2 monolayers of gold at a substrate temperature of 770-870 K and subsequent annealing to 1070~K. Low-energy electron diffraction (LEED) then showed a sharp $5 \times 1$ diffraction pattern characteristic for the Si(557)-Au surface. For the STM and STS measurements performed in a commercial LT-STM (Omicron Nanotechnology, Germany) at $T$ = 300 K and 77 K, clean and sharp tungsten tunneling tips were prepared by in-situ heating, sputtering, and field emission.
STM images were recorded by scanning the tip parallel to the chain direction in order to reduce the action of the feedback loop to obtain the highest spatial resolution.

\section{Results and Discussion}
\begin{figure}[h!]
\includegraphics[width=\columnwidth]{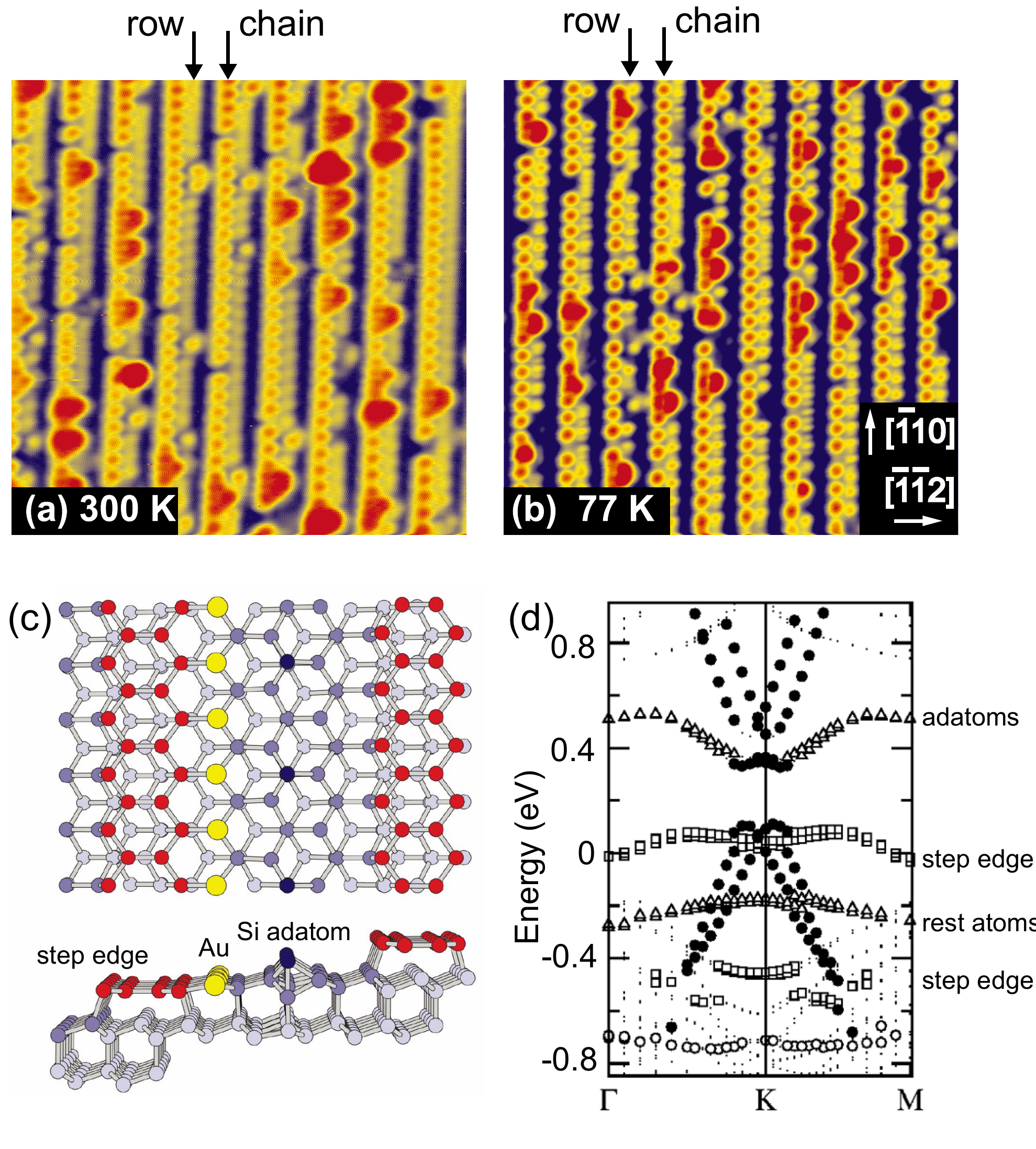}
\caption[]{STM images (20 nm $\times$ 20 nm) of Si(557)-Au taken at a sample voltage $U = +1$ V (tunneling current $I$ = 0.1 nA) at (a) room temperature and at (b) $T$ = 77 K. The terrace steps of the vicinal surface are ascending from left to right. (c) Structure model for the Si(557)-Au surface. From Crain \textit{et al.} \cite{Crain04}. (d) Calculated band structure (including spin-orbit interaction) for the Si(557)-Au surface. Filled circles indicate bands arising from Au atoms at substitutional Si positions in the middle of the terrace. From S\'{a}nchez-Portal \textit{et al.} \cite{Sanchez04}.} 
\label{fig1}
\end{figure}

Typical STM images of the unoccupied electronic states at RT and LT (Fig. \ref{fig1}) show the characteristic surface structure of Si(557)-Au with randomly distributed larger protrusions presumably due to excess Si expelled from the top Si layer during Au-chain formation \cite{Kirakosian03}. The two prominent 1D features are denoted as ``chain'' of localized protrusions) and ``row'' \cite{Schoeck06}. According to a structure model of the Si(557)-Au surface derived from previous x-ray scattering experiments \cite{Robinson02} (Fig. \ref{fig1}c) and corroborated by {\it ab-initio} calculations \cite{Sanchez02} (Fig. \ref{fig1}d), the row is attributed to Si atoms forming the step edge of the vicinal surface while ``chain'' is attributed to a chain of Si adatoms on every terrace. 

\begin{figure}
\includegraphics[width=\columnwidth,clip=]{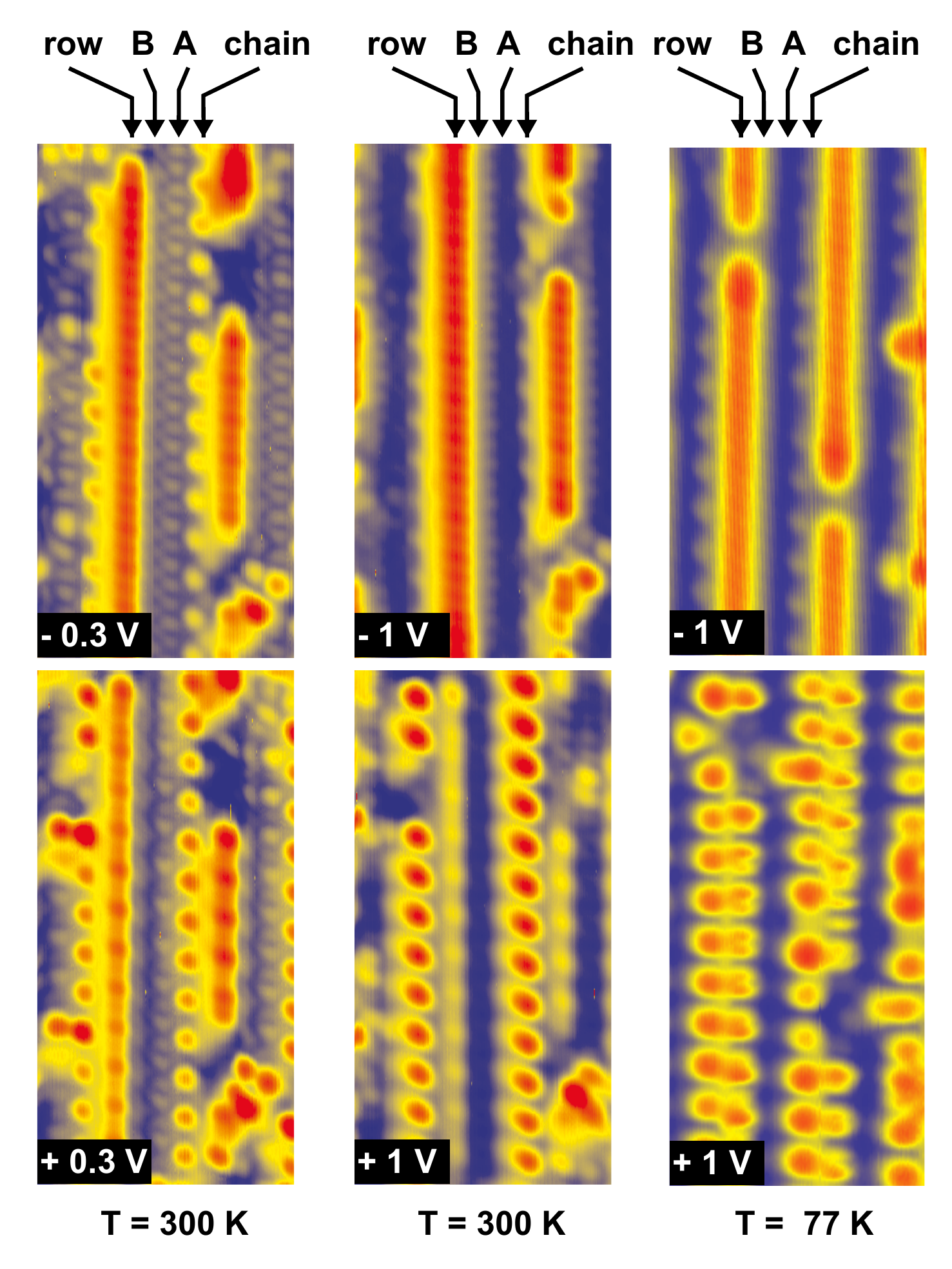}
\caption[]{STM images (10 nm $\times$ 5 nm) of Si(557)-Au taken at different voltages $U$ at room temperature  and at $T$ = 77 K. The tunneling current was $I$= 0.05 nA for all room temperature measurements. At $T$ = 77 K, tunneling currents of $I$ = 0.01 nA ($U = -1$\,V) and $I$ = 0.03 nA ($U = + 1$\,V) were used.} 
\label{fig2}
\end{figure}

Figure \ref{fig2} shows voltage-dependent STM images taken at RT and LT. Besides the localized protrusions of the adatom chain, the periodic modulation of the intensity along the step edge (``row'') is clearly resolved for positive voltages at high and low temperatures, as observed earlier \cite{Yeom05,Sauter07}. In more than 80 \% of our STM images, the phase of this periodic modulation on the row changes by half a period when reversing the tunneling bias. This is due to the alternating LDOS of occupied (unoccupied) states located at the upper (lower) step-edge atoms of the buckled step edge \cite{Sanchez04}. Moreover, the apparent length of the row segments depends on the voltage polarity. For negative bias $-1$\,V each segment appears longer at each end by one Si-Si interatomic distance ($a$ = 0.384 nm along [\=110]) compared to positive bias $+ 1$\,V, similar to Ref. \onlinecite{Krawiec06}. This is presumably due to the presence of electronic ``end states'' previously reported for Si(553)-Au \cite{Crain05}. These end states are still observed at $T$ = 77 K indicating that the electron dynamics within a single row is not changed, contrary to what one might expect for a Peierls transition. 

The appearance of the periodically arranged localized protrusions attributed to a chain of Si adatoms is almost independent of the tunneling voltage. Two additional weak features denoted by A and B are found for small voltages $U = +0.3$~V and $U = -0.3$ V between row and chain. This observation demonstrates the high resolution obtained in the STM images, and is of particular interest because according to the structure model it is in this region where the Au atoms are located \cite{Robinson02}. Another STM study has resolved features similar to A and B which have been discussed in agreement with the structural model \cite{Han2009,Sanchez04}. 

\begin{figure}
\includegraphics[width=\columnwidth]{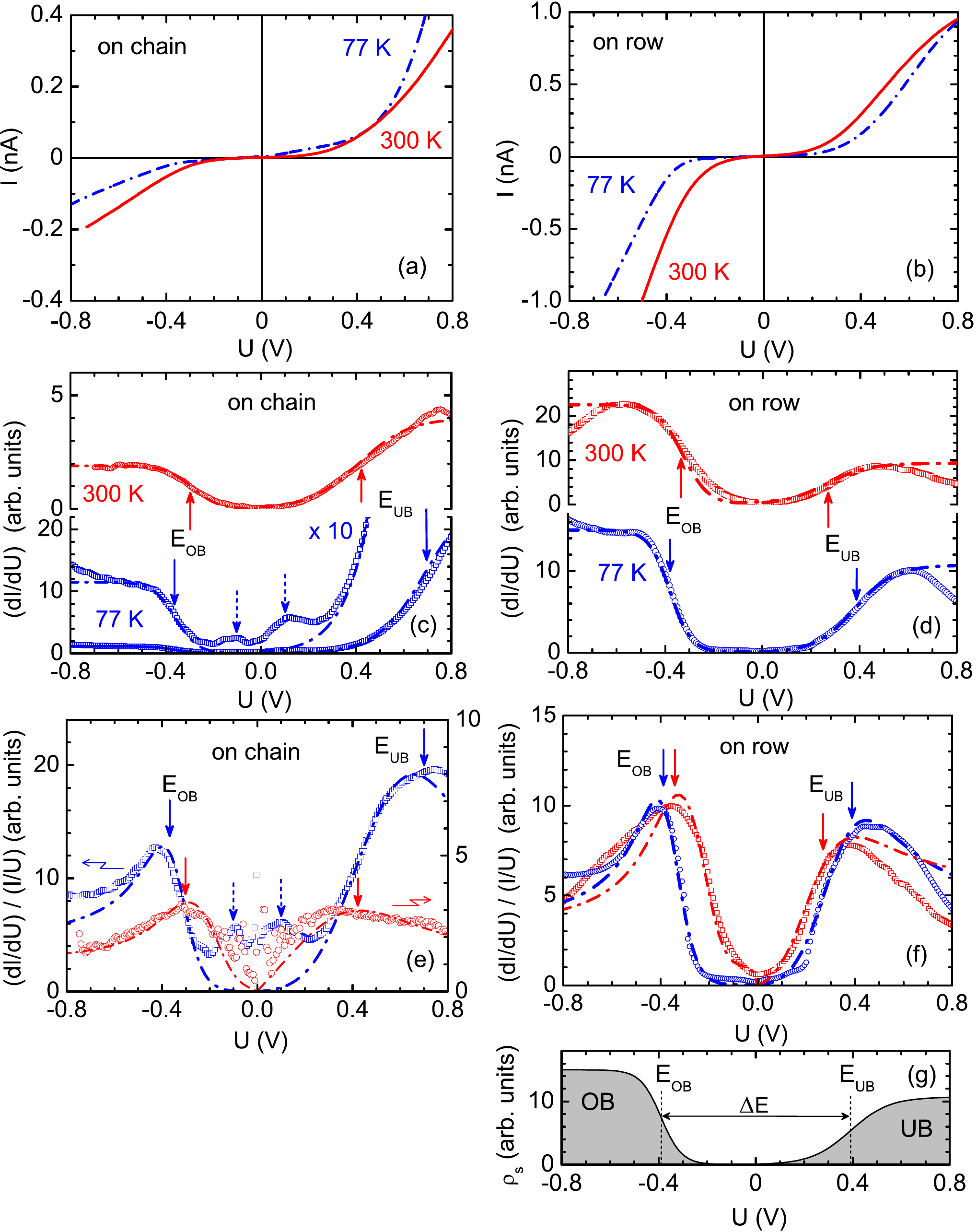}
\caption[]{(a,b) $I(U)$ characteristics taken at features labeled ``chain'' and ``row'' at room temperature (solid line) and at $T$ = 77 K (broken line). (c,d) Tunneling conductance $dI/dU$ obtained from the $I(U)$ curves (open symbols). Dashed arrows indicate two contributions on ``chain'' which occur at $U = \pm 0.1$ V for $T$ = 77 K. Dash-dotted lines show the calculations according to Eqs. \ref{eq1} and \ref{eq2}, see text for details. Solid arrows denote the positions of the OB and UB LDOS edges $E_{OB}$ and $E_{UB}$, respectively, obtained from the calculations. (e,f) Normalized tunneling spectra $(dI/dU)/(I/U)$. (g) Schematic of the LDOS (Eq. \ref{eq2}) used for the calculation of $dI/dU$ (see  Table \ref{table1} for the parameters of ``row'' at 77 K).} 
\label{fig3}
\end{figure}

In the following, we will compare the atomically resolved $I(U)$ characteristics and tunneling spectra taken at positions on the chain and on the row (Fig. \ref{fig3}). First, we carefully checked that the measured $I(U)$ characteristics [Fig. \ref{fig3}(a,b)] are in agreement with our voltage-dependent images. The resolution of the RT data has been significantly enhanced in comparison to our previous results \cite{Schoeck06}. The tunneling conductance $dI/dU$ was numerically derived from an average of 20-30 $I(U)$ curves. The $dI/dU$ spectra shown in Fig. \ref{fig3}(c,d) exhibit a gapped LDOS with almost zero intensity around the Fermi level ($U$ = 0) except for the chain at $T$ = 77 K. The latter shows two distinct peaks located at $U = \pm 0.1$ V that were reproducibly observed with several different STM tips, see dashed arrows in Fig. \ref{fig3}(c). Note that these features are already visible in the $I-U$ curve [Fig. \ref{fig3}(a)]. The spectrum on the row [Fig. \ref{fig3}(d)] is similar to $dI/dU$ spectra taken at RT on the step edge reported by Kang \textit{et al.} \cite{Kang2009}. Spectra taken in the region {\it between} the row and the chain, i.e., near A and B (see Fig. \ref{fig2} at $U = \pm$ 0.3 \,V) are  similar to spectra taken on the row. 

For comparison with previously published data Figs. \ref{fig3}(e,f) show the normalized differential conductance $(dI/dU)/(I/U) \approx d{\rm ln}I/d{\rm ln}U$ which compensates for the exponential dependence on voltage and is independent of the tip-sampe separation \cite{Feenstra,Zhang}. The $(dI/dU)/(I/U)$ data taken on the chain at $T$ = 300 K [Fig. \ref{fig3}(e)] scatter for small bias voltage due to the very low current, cf. Fig. \ref{fig3}(a). In contrast, $(dI/dU)/(I/U)$ data on the row [Fig. \ref{fig3}(f)] do not show this behavior possibly to a weak but finite current at low bias, cf. Fig. \ref{fig3}(b). The nonzero conductance on the row might explain the observed dependence of the $dI/dU$ \,maps for small tunneling voltages reported by Han \textit{et al.} \cite{Han2009} suggesting metallicity of the step-edge atoms at room temperature. The normalized spectra on the row are in good agreement with data reported by Yeom \textit{et al.} \cite{Yeom05}. However, in that case, the data were interpreted as indication for an energy gap occurring at LT which is strongly reduced to 0.08 - 0.1 eV at RT \cite{Yeom05}. 

In the following we investigate these points in more detail by extracting the LDOS from the differential tunneling conductance $dI/dU$. In order to precisely determine the gap width and thermal broadening from the tunneling spectra, we have used the temperature-dependent differential tunneling conductance according to the elastic-tunneling theory \cite{Bardeen,Lang}:
 
\begin{equation}
\frac{dI}{dU} (U, T) = \int^{+\infty}_{-\infty}D(E, U)\rho_s(E)\rho_t\left[\frac{d}{dU}f(E-eU,T)\right]dE
\label{eq1}
\end{equation}

where $\rho_s(E)$ and $\rho_t$ are the LDOS of the sample and of the STM tip, respectively. The latter is assumed to be featureless. $f(E-eU,T)$ is the Fermi-Dirac distribution function with the Fermi energy corresponding to $E = 0$. The tunneling transmission factor $D(E,U)$ is a slowly varying function of $E$ and $U$ in the voltage range $\left| E-U \right| \ll \phi$ considered here ($\phi$: averaged work function of tip and sample). Since we are only interested in the positions and the broadening of the occupied band (OB) and unoccupied band (UB) edges, $\rho_s (E)$ was simply modelled by two constants $\rho_{OB}$ and $\rho_{UB}$ up to 
energies $E_{OB}$ and $E_{UB}$ of the OB and UB edges, respectively, see Fig. \ref{fig3}(g) for an example: 

\begin{equation}
\rho_s(E) = \frac{\rho_{OB}}{{\rm exp}(\frac{E-E_{OB}}{\Gamma_{OB}})+1} + \frac{\rho_{UB}}{{\rm exp}(\frac{E_{UB}-E}{\Gamma_{UB}})+1}
\label{eq2}
\end{equation}

The widths of the band edges, $\Gamma_{OB}$ and $\Gamma_{UB}$, were adjusted to obtain the best fit of Eq. \ref{eq1} to the data at $T$ = 77 K and were considered to be temperature independent.

Figs. \ref{fig3}(c,d) show the measured $dI/dU$ data together with the differential tunneling conductance calculated according to Eqs. \ref{eq1} and \ref{eq2} with band-edge energies $E_{OB}$ and $E_{UB}$ indicated by arrows. Due to the nonlinear $I(U)$ characteristics the band edges give rise to peaks in the normalized spectra $(dI/dU)/(I/U)$ shown in Fig. \ref{fig3}(e,f) \cite{Zhang}. The parameters obtained from the calculations are given in Table \ref{table1}. The good agreement between data and fits demonstrates the consistency of our data analysis. The effect of variations in $D(E,U)$ on $dI/dU$ was checked for reasonable values of $\phi$ and for several tip-sample distances. These variations were found to change the energy gap $\Delta E= E_{UB} - E_{OB}$ by less than 0.02 eV. Evidently, the broadenings $\Gamma_{OB}$ and $\Gamma_{UB}$, are much larger than the thermal broadening of the Fermi-Dirac function. A possible origin of the widths of the band edges might be found in the finite length of the one-dimensional segments, typically in the range of 10 - 15 nm. With the Heisenberg uncertainty relation $\Delta x \Delta p \approx \hbar/2$ we obtain for the energy uncertainty $\Delta E = \hbar k_{\rm F}\Delta p/m =\hbar^2 k_{\rm F}/2m\Delta x \approx$\, 0.06 eV with $\Delta x$\,= 10 nm, $k_{\rm F} = 2\pi/a$, and electron mass $m$, in order of magnitude agreement with the experimental data. $\Gamma_{UB}$ is systematically larger than $\Gamma_{OB}$, in line with this qualitative argument. This motivates future investigations on the dependence of the LDOS on the length of the  one-dimensional segments.

Table \ref{table1} shows that for each temperature the energy gap $\Delta E$ is wider on the chain than on the row. After heating from 77 K to 300 K it decreases by 0.35 eV on the  chain and by 0.17 eV on the row. Hence, the LDOS energy gap decreases considerably on both characteristic features, row \textit{and} chain, of the Si(557)-Au surface and not solely on the row attributed to the step edge. Note that the gap reduction with increasing $T$ is even larger for the chain than for the row. This thermal effect on the LDOS cannot therefore straightforwardly be attributed to a metal-insulator transition (MIT) occurring on the row as suggested by band-structure calculations. The existence of an MIT was inferred from photoemission data acquired in the first Brillouin zone \cite{Ahn03}, whereas an MIT was excluded on the basis of high-intensity ARPES using synchrotron radiation \cite{Kim2009}. In the latter experiments, a satellite peak observed in the Au core-level spectrum was attributed to the excitation of a one-dimensional plasmon in the Au-Si metallic chains. 

When comparing STM/STS results to the electronic band structure one has to keep in mind that (i) effects due to electron-electron interaction are usually not taken into account in the tunneling model and (ii) STS is sensitive to electronic states located at or close to the $\bar{\Gamma}$ point of the surface Brillouin zone, i.e., at $k_{\parallel} =0$, due to the energy-dependent transmission factor $D(E, U)$. Away from the $\bar{\Gamma}$ point only flat bands with a high LDOS may contribute. An inverse photoemission study on Si(557)-Au \cite{Lipton} shows two closely separated bands crossing the Fermi level at $k_{\parallel} \approx 0.5$ \AA$^{-1}$, in agreement with earlier photoemission results \cite{Ahn03,Altmann}. At the $\bar{\Gamma}$ point ($k_{\parallel} = 0$), a weak contribution was observed at approximately $0.5$\,eV above the Fermi level. This energy agrees roughly with the position of the maxima at $+0.6$\, V (LT) and $+0.4$\, V (RT) in our STS spectra taken on the row, see Fig. \ref{fig3}(d) and Ref. \onlinecite{Schoeck06}. However, in inverse photoemission a state with even higher intensity was found on the bare Si(557) surface at $+0.5$\, eV as well and was identified as a Si(111)-(7$\times$7) U$_1$ adatom state \cite{Lipton}. This makes the comparison of the STS and inverse photoemission data for this particular state rather ambiguous. 
\begin{table}
\caption{Parameters obtained from the tunneling conductances (Eqs. \ref{eq1} and \ref{eq2}),  
see Fig. \ref{fig3} and text for details.}
\begin{ruledtabular}
\begin{tabular}{c|rr|rr} 
\multicolumn{1}{c|}{Feature}&\multicolumn{2}{c|}{chain}& \multicolumn{2}{c}{row}\\
\multicolumn{1}{c|}{$T$ (K)}&\multicolumn{1}{c}{77}& \multicolumn{1}{c|}{300}& \multicolumn{1}{c}{77} & \multicolumn{1}{c}{300}\\
\hline
$\Gamma_{OB}$ (eV)  &$0.045$&$0.045$ &$0.042$&$0.042$\\
$\Gamma_{UB}$ (eV)  &$0.100$&$0.100$&$0.080$&$0.080$\\
$E_{OB}$ (eV) & $-0.37$ & $-0.30$ & $-0.39$ & $-0.34$\\
$E_{UB}$ (eV) & $+0.70$ & $+0.42$ & $+0.39$ & $+0.27$\\
$\Delta E$ (eV)  & $ 1.07$ & $0.72$ & $0.78$ & $0.61$\\
\end{tabular}
\end{ruledtabular}
\label{table1}
\end{table}

When comparing our STS data with surface band-structure calculations of Si(557)-Au \cite{Sanchez04} shown in Fig. \ref{fig1}(d) we find partial agreement. From the calculations, a bulk band gap of 0.92 eV and a surface band gap of approximately 0.7 eV between the occupied bands of the ``rest atoms'' - located at positions A or B between the adatoms and the step edge - and the unoccupied bands of the adatom chain are estimated at the $\bar{\Gamma}$ point. However, two-photon photoemission experiments employing a femtosecond pump-probe technique found an experimental bulk band gap of 1.16 eV \cite{Ruegheimer} between the valence band maximum at $\approx -0.2$\, eV and the conduction band minimum (CBM) at $\approx 0.96$\, eV ($T$ = 90 K). Since the bulk band gap of semiconductors is often underestimated in DFT calculations, a down shift of the experimental \textit{unoccupied} bands by 0.24 eV was suggested in order to line up the calculated and experimental CBM \cite{Ruegheimer}. Taking this into account, the weakly dispersing unoccupied band between $+0.51$\,eV ($\bar{\Gamma}$ point) and $+0.32$\,V ($\bar{\rm K}$ point) arising from adatoms (``chain'') [Fig. \ref{fig1}(d)] should lead to an increasing tunneling conductance between $+0.75$\, V and $+0.56$\, V. This is in agreement with the tunneling spectra taken on the chain at $T$ = 77 K, cf. Fig. \ref{fig3}(c,e). Furthermore, a similarly inferred surface band gap of $\approx 1$\, eV at the chain is in fair agreement with our result $\Delta E= 1.07$ eV, see Table \ref{table1}. 

The weak maximum in the tunneling spectra of the row at $\approx -0.5$\,V [Fig. \ref{fig3}(d)] can be attributed to the step edge in agreement with the band-structure calculations \cite{Sanchez04}, whereas the maximum at $\approx\, +0.6$\,V cannot be ascribed to a theoretically predicted band for the step edge. Moreover, we do not observe contributions close to the Fermi level which could be attributed to the predicted flat surface-state band arising from the step-edge atoms \cite{Sanchez04}, see Fig. \ref{fig1}(d). This is in agreement with the fact that these step-edge states close to the Fermi level have never been observed in photoemission. Although low-energy contributions are detected at $\pm 0.1$ V on the chain [Fig. \ref{fig3}(c), $T$ = 77 K], they are not found on the row, i.e, on the step edge. 

In two-photon photoemission experiments, three new states near the $\bar{\Gamma}$\, point that cannot be explained by existing band-structure calculations \cite{Sanchez04} were found at $T$ = 90 K \cite{Ruegheimer}: a state at $\approx -0.4$\, eV with lifetime $\tau < 10$\, fs attributed to an image-potential resonance, a state at $\approx +0.5$\, eV that was only observed after a delay time of $\approx 100$\, fs, and a surface state at $\approx +0.9$\, eV close to the CBM. Although these positions are compatible with energies $eU$ where maxima are observed in the tunneling conductance at $T$ = 77 K, cf. Figs. \ref{fig3}(c,d), it is presently not clear if these states will contribute to the tunneling current. 
 
We also observe an LDOS variation along the step edge due to occupied (unoccupied) states located at the upper (lower) step-edge atoms as predicted. Furthermore, features A and B observed in the spatial region where the Au chain is supposedly located \cite{Robinson02,Crain04}, are resolved in scans parallel to the 1D features for small voltages only, where in the constant-current mode the tip is close to the surface. This suggests that the electron wave functions of the Au atoms embedded in the Si surface do not extend far into the vacuum contrary to what might be expected for the $s$ and $p$ character of the Au and Si-Au bonds \cite{Sanchez04}. However, these bands have a strong dispersion for $k_{\parallel} \gg 0$ and cross the Fermi level near the \=K point of the surface Brillouin zone. They therefore do not contribute significantly to the tunneling current.

Hence, although some relevant energies can be estimated from our experimental data and compare nicely with theory, we do not find complete agreement with the band-structure calculations.

\section{Conclusion}
STM images and atomically resolved tunneling spectra were measured on the Si(557)-Au surface at 77 K and 300 K. Comparison of the $T$-dependent data shows that the energy gap $\Delta E$ in the tunneling spectra observed at low temperatures becomes considerably smaller at room temperature on {\it both} characteristic protrusions of this surface. Our data support a change of the band edge positions with temperature and indicate a $T$-induced modification of the LDOS of both features. Furthermore, the STM images do not show evidence for a  modification of the structural units or a change in the periodicities, in agreement with previous investigations \cite{Sauter07,Han2009}. Hence, the metal-insulator transition inferred from photoemission data \cite{Ahn03,Yeom05} cannot solely involve the step edge as suggested by band-structure calculations. A complete understanding of the mutual relationship between structural and electronic properties of this surface strongly requires a refinement of the band structure calculations for the Si(557)-Au surface. Our new data suggest that more attention should be paid to finite-size effects on the one-dimensional segments characteristic for this surface.

\begin{acknowledgments}
We thank H. Stalzer for technical help. This work was supported by the Deutsche Forschungsgemeinschaft through the DFG Center for Functional Nanostructures (CFN).
\end{acknowledgments}

\end{document}